\documentclass[12pt]{iopart}

\usepackage{xcolor}
\usepackage{graphicx}
\usepackage{subcaption}
\usepackage{booktabs}
\usepackage{hyperref}
\usepackage{nicefrac}
\usepackage{soul}

\newcommand{\strev}[1]{}
\newcommand{\rev}[1]{#1}
\newcommand{\revcom}[1]{}
\newcommand{\revd}[1]{#1}

\expandafter\let\csname equation*\endcsname\relax
\expandafter\let\csname endequation*\endcsname\relax

\usepackage{amsmath}
\usepackage{amssymb}

\begin{document}

\title[Acc. NR Simulations with New Multistep fourth-order RK Methods]{Accelerating Numerical Relativity Simulations with New Multistep Fourth-Order Runge-Kutta Methods}

\author{Lucas Timotheo Sanches\footnote{Corresponding author}}
\address{Center for Computation and Technology -- Louisiana State University, 888 S. Stadium Dr., Baton Rouge, LA, USA}
\ead{lsanches@lsu.edu}
\vspace{10pt}

\author{Steven Robert Brandt}
\address{Center for Computation and Technology -- Louisiana State University, Baton Rouge, LA, USA}
\ead{sbrandt@cct.lsu.edu}
\vspace{10pt}

\author{Jay Kalinani}
\address{Center for Computational Relativity and Gravitation School of Mathematical Sciences --- Rochester Institute of Technology, Rochester, NY, USA}
\ead{jaykalinani@gmail.com}
\vspace{10pt}

\author{Liwei Ji}
\address{Center for Computational Relativity and Gravitation School of Mathematical Sciences --- Rochester Institute of Technology, Rochester, NY, USA}
\ead{ljsma@rit.edu}
\vspace{10pt}

\author{Erik Schnetter}
\address{Perimeter Institute --- Waterloo, ON, Canada}
\address{Department of Physics and Astronomy --- University of Waterloo, Waterloo, ON, Canada}
\ead{eschnetter@perimeterinstitute.ca}
\vspace{10pt}

\begin{indented}
  \item[]June 2025
\end{indented}

\begin{abstract}
Many HPC applications that solve differential equations rely on the Runge-Kutta  family of methods for \textit{time integration}. Among these methods, the fourth-order accurate RK4 scheme is especially popular. This time integration scheme requires applications to evaluate four intermediate stages to take one time step. Depending on the complexity of the problem being solved, the evaluation of these intermediate stages can be computationally expensive.

In this paper we develop explicit \strev{third and} fourth-order accurate Multistep Runge-Kutta (MSRK) methods. The advantage of such methods is that they re-use data from previous time steps, thus requiring fewer  intermediate stage evaluations and potentially speeding up applications. We  outline a procedure to obtain and tune the method's coefficients by adjusting their stability regions in an attempt to maximize the size that a time step can take. We validate and evaluate our new methods in the context of Numerical Relativity applications using the \texttt{EinsteinToolkit}. We believe, however, that these methods and results should generalize to other applications using explicit Runge-Kutta methods.
\end{abstract}
\noindent{\textit{Accepted for publication in Classical and Quantum Gracity}}

\vspace{2pc}
\noindent{\it Keywords}: Multistep Runge-Kutta Methods, Numerical Relativity, Binary Black Hole Simulations, \texttt{EinsteinToolkit}, \texttt{CarpetX}, GPUs.
%
%
%
%

\section{Introduction} \label{sec:intro}

As a result of LIGO's Nobel Prize-winning
discovery of gravitational waves emitted by merging black holes~\cite{Abbott:2016blz},
as well as the landmark observation of colliding neutron stars in both the gravitational
and electromagnetic radiation 
spectra~\cite{TheLIGOScientific:2017qsa},
a new era of gravitational wave physics and multimessenger astrophysics has begun.
However, these exciting new discoveries rely on theoretical predictions from the numerical relativity and computational astrophysics communities. Because the parameter space for these numerical calculations is large and single detections may need to compare against many of them, it is desirable to make these relativistic codes as fast as possible.

A key ingredient of such simulations is known as the \textit{time-integrator}, a mathematical algorithm for approximating the next state of a system given its current (and possibly past) states. The time-integrator can be chosen from among what are known as \textit{implicit} or \textit{explicit} methods. Among the family of explicit methods (what we are considering here), the Runge-Kutta (RK) methods are the most commonly employed schemes. This is particularly true in Numerical Relativity (NR). This popularity is due to their stability properties, simplicity, and because they are well understood mathematically.

These methods evaluate the time derivatives of a system multiple times within the computation of a time step update. This set of evaluations is collectively known as \textit{stages}. When combined, these stages provide stable, efficient, and accurate methods to evolve systems of equations forward in time.

Alternative explicit time-integration algorithms are offered by the Linear Multistep (LM) or Adams-Bashforth (AB) family of methods. These methods use past time steps and the current time step (stage) to extrapolate a future time step. Computationally speaking, this approach increases the memory requirements of the algorithms but reduces their operational complexity when compared to RK methods, since there is only a single stage that needs to be computed. \revd{However, apart from the third-order Adams-Bashforth method (AB3)~\cite{KIDDER201784}, these methods require time steps that are too short} for the applications of interest in NR, nullifying any computational advantages they may obtain by computing fewer stages.

To characterize the computational efficiency of a time integration scheme, it is common to compute the ratio between the maximum time step a method can take (without rendering the evolution unstable) and some measure of the amount of computational work the method performs in advancing the solution. For an explicit RK scheme, the amount of work is taken to be that of the number of stages the method requires. This is because evaluating a single stage is likely to be the most computationally intensive task an application may have (this is certainly true in NR).

We note that evaluation of the time derivative on the current step counts as both a ``step'' and a ``stage'' in common terminology. Thus, a standard RK4 method uses one step but four stages, while a fourth-order LM method uses a single stage but four steps.

What many researchers, particularly in NR, appear to be unaware of, however, is that both RK and LM methods are subsets of the well-studied broad class of \textit{General Linear Methods}, which can combine steps and stages in different ways. Refs.~\cite{jackiewicz_glm_book, butcher_2006} provide a comprehensive discussion on this topic. Depending on the number of stages and the various coefficients used (e.g. two steps and three stages, or three steps and two stages for a fourth-order method), one can arrive at algorithms that are more efficient than the single-step Runge-Kutta methods. In this paper, we will specifically look at \textit{Multistep Runge-Kutta} (MSRK) methods, sometimes called ``Accelerated Runge-Kutta'' methods in the literature. 

The main contributions of this paper are as follows: (1) three new fourth-order MSRK methods, (2) the techniques by which we derived these methods, and (3) the ``dense output'' formulas for our new methods, which can be used to extrapolate fields to obtain values between time steps. We test and validate our methods by performing a variety of tests and comparisons in NR simulations. Our most important result, which compares the performance of our new methods to that of RK4 and \textit{low storage methods} is summarized in Tab.~\ref{tab:result_summary}. To our knowledge, our work is the first successful attempt to construct and utilize these methods for Numerical Relativity applications, in particular, for binary black hole collision simulations.

\section{Constructing Multistep Runge-Kutta methods and related works}\label{sec:msrk}

In this section, we will briefly review the mathematical theory of Runge-Kutta, Linear Multistep methods and their stability analyses.

We will derive and present two novel fourth-order accurate explicit MSRK methods: two using 2 time steps and 3 intermediate stages, which we name \texttt{RK4-2(1)} and \texttt{RK4-2(2)}; and one using 3 time steps and 2 intermediate stages, which we name \texttt{RK4-3}. 

The methods we propose fall under the broad category of General Linear Methods. These methods contain RK methods, LM methods, and combinations of these methods. The mathematical theory behind general linear methods and MSRK methods is very broad and well developed. It seems, however, relatively unknown in NR. In our work, we will only present the necessary ingredients for the adequate understanding of our techniques, but we encourage interested readers to seek Refs.~\cite{jackiewicz_glm_book, butcher_2006, suli_mayers_2003, leveque2007, butcher_2008} for more information.

Our RK4-2 methods are classified in the mathematical literature as Two-Step Runge-Kutta (TSRK) methods as they use 2 time steps and 3 stages. References~\cite{jackiewicz_tracogna_tsrk_original,Jackiewicz_tsrk_2} are the seminal works for this category of methods and provide detailed mathematical studies and explicit examples of both implicit and explicit TSRK methods.

As far as the authors are aware, the number of scientific codes that make use of MSRK methods appears to be small (or null in the case of NR). Our survey in the literature revealed only one, given in Ref.~\cite{openfoam_accelerated_rk}, where the authors implement a third order two-step RK method first proposed in Ref.~\cite{accelerated_rk} in the \texttt{OpenFOAM} CFD infrastructure. In this work, the authors refer to these methods as \textit{Accelerated} Runge-Kutta methods.

\rev{In our work, we have chosen to focus on a particular class of 2 and 3 step methods which we will detail in Sec.~\ref{sec:hybrid_method}, namely those methods which re-use previous values of the right-hand side (RHS) of the evolution equations.

We chose this limitation for several reasons. Firstly, to reduce the search space. Secondly, to minimize the computation of linear combinations and storage overheads. Finally, we wanted to work with something similar in form to the standard RK4 algorithm. In future works, we may, however, consider looking at wider classes of methods.}

\subsection{Runge-Kutta and Linear Multistep Methods}\label{subsec:rk_and_lmm}

To solve the PDEs arising in NR (and many other fields), the Method of Lines is customarily employed~\cite{mol1,mol2,mol3}. The idea of the method is to semi-discretize the PDEs in space to obtain a system of coupled ODEs that can then, in turn, be solved with standard ODE solvers. We are, therefore, interested in methods to solve the Cauchy Problem:
\begin{equation}
    \frac{\mathrm{d} y(t) }{\mathrm{d}t} = f(t,y),\quad y(t_0) = y_0 
    \label{eq:standard_ivp}
\end{equation}
where $t_0$ is an arbitrary initial value for $t$, and $y_0$ an arbitrary initial condition.

The first step in solving Eq.~\eqref{eq:standard_ivp} is to discretize $t$ into a set of \textit{time steps} $t_i$, where each step is given by
\begin{equation}
    t_i = t_0 + n h,
    \label{eq:time_steps}
\end{equation}
where $n$ is an integer and $h$ is known as the \textit{time step size}. Given these definitions, we introduce the notation
\begin{equation}
    y_n \equiv y(t_n).
    \label{eq:notation_for_yn}
\end{equation}

In order to evolve the Cauchy Problem described by Eq.~\eqref{eq:standard_ivp} from state $y_n$ at a time $t_n$ to state $y_{n+1}$ at time $t_n + h$, RK methods are often employed. They compute the next time step in the numerical evolution by calculating several intermediate \textit{stages}. Stages are additional evaluations of $f(t,y)$, which we refer to as the RHS of Eq.~\eqref{eq:standard_ivp}, at intermediate points between $t_n$ and $t_n + h$, which are later combined in an update formula. A general explicit $s$-stage RK method can be written as~\cite{leveque2007,butcher_2008}
\begin{equation}
    y_{n+1} = y_n + h \sum_{i=1}^{s} b_i k_i,
    \label{eq:general_rk_method}
\end{equation}
where $k_s$ are the stages of the method, given by
\begin{equation}
    k_s = f(t_n + c_s h, Y_s)
    \label{eq:rk_stage}
\end{equation}
and the $Y_s$ are given by
\begin{equation}
    Y_s = y_n + h \sum_{i = 1}^{s-1}a_{si}k_i.
\end{equation}

LM methods, on the other hand, compute $y_{n+1}$ by employing a linear combination of $y_n$ and various previous \textit{steps}. A general $r$ step LM is given by~\cite{leveque2007, butcher_2008}.
\begin{equation}
    \sum_{j=0}^{r} \alpha_j y_{n+j} = h \sum_{j=0}^{r} \beta_j f(t_{n+ j}, y_{n+j}),
    \label{eq:general_lmm}
\end{equation}
Choosing $\beta_r = 0$ is required for explicit methods.
We note that the evaluation of \rev{$y_{r-1}$}\strev{$\beta_{r-1}$} counts as both a step and a stage in the above terminology, as it involves an explicit calculation of $f(t, y)$~(See Eq.\eqref{eq:standard_ivp}). The values of $\beta_j$ where $j < r-1$ can be implemented as saved values of previous evaluations of $\beta_{r-1}$.

\subsection{Region of absolute stability of ODE solvers}\label{subsec:stability_of_ODE_solvers}

To analyze our new methods, we will plot their absolute stability region and compare it against that of the standard RK4 method.

\revcom{Discussion of stability has been removed}

Our visualization algorithm is based on that of Ref.~\cite{leveque2007}. We first apply the equation for $y_{n+1}$ (In the case of RK methods and LM methods Eqs.~\eqref{eq:general_rk_method} and \eqref{eq:general_lmm}, respectively) to the model problem \rev{$y^\prime(t) = \lambda y(t)$}. We then introduce the quantity $z\equiv\lambda h$, allowing for imaginary values of $\lambda$ (and thus of $z$). After that, we replace each resulting $y_{n+i}$ by the complex variable $\zeta^{n+i}$ (the superscript here indicates powers of $\zeta$ and not an index), constructing a polynomial $\rho(\zeta;\;z)$ in $\zeta$ known as the \textit{stability polynomial} of the method. The ASR, then, is the locus of all complex points $z$ where the absolute values of all roots of the stability polynomial are smaller than one, that is,
\begin{equation}
  \left| r_n(z) \right| < 1 \; \forall \; n.
  \label{eq:asr_def}
\end{equation}
\revd{Where $r_n(z)$ is the $n$-th root of $\rho(\zeta;\;z)$. Equation \eqref{eq:asr_def} is known in the literature as the \textit{root condition} for absolute stability.}

The shape of the ASR of a given time integrator is of great practical importance. When discretizing a PDE system via Method Of Lines, it can be shown that for a time integrator to successfully solve the resulting ODEs, the semi-discrete system's characteristic eigenvalues need to fall within the method's ASR. By performing spatial and temporal discretizations of the scalar wave equation one can see that the system has purely imaginary characteristic eigenvalues which impose a limitation on the type of time integrators that can be used. We note that the wave equation shares important mathematical similarities with the evolution equations used in General Relativity, and so it has been commonly used to study the stability of numerical schemes. See Ref.~\cite{peter_wave_eq} for a detailed description of the above.

Generally speaking, for systems that contain purely imaginary eigenvalues it is desirable that the ASR contain a large portion of the imaginary axis. In practice, this means that the larger the value of the \textit{intercept} of the ASR with the imaginary axis, the larger the potential time step one may take during the numerical evolution.
\strev{Looking again at Fig.}~\ref{fig:stability_compare}\strev{, we can see that both the Euler scheme and the 2-step AB scheme are unconditionally unstable for such systems.}

\strev{Alternatively, RK4 and Bu4-2 do cover part of the imaginary axis, and are, therefore, stable for this problem for some choice of time step and spatial discretization.}

\subsection{New families of multistep RK time integrators}\label{sec:hybrid_method}

Before continuing, it is important for us to recall and appreciate the significance of the so called \textit{Courant-Friedrichs-Lewy} number of a given numerical scheme. Even though time integrators, when used for solving ODEs, require only the selection of a suitable time step, when solving PDEs via the method of lines the spatial discretization size becomes important for stability. The ratio between the time step size and spatial step size of a scheme is known as the Courant-Friedrichs-Lewy number (CFL) for the scheme. A (necessary but not sufficient) condition for the stability of the scheme is that the CFL is below a certain threshold, depending on the system being solved and the discretization schemes being employed in the solution~\cite{cfl_1928}. We can characterize the computational efficiency of the numerical scheme by computing the \textit{Effective CFL number} (ECF), which is the ratio between the maximum CFL possible for a stable evolution of a system and the number of intermediate time integrator stages being employed. Once again, this measure assumes that the evaluation of the RHS of an ODE system (in this case obtained after spatial discretization) is the most computationally intensive task in the numerical evolution.

Alternative RK methods have been proposed to alleviate computational costs and improve the stability of numerical schemes solving PDEs. In particular, we'd like to call attention to the family of so-called \textit{Low Storage} methods, introduced by Williamson in Ref.~\cite{williamson_198048} and studied and improved in detail by Ketcheson in Refs.~\cite{Ketcheson_2010, Ketcheson_2012, Ketcheson_2013} and related works. These methods allow for larger CFL conditions with the trade-off that they require more intermediate stages to be computed. They have found successful use in several applications, but of particular interest to us is their use in \texttt{AthenaK}~\cite{athenak}, a GPU accelerated framework for NR simulations.

Full numerical evolution of Einstein's equations requires \textit{gauge conditions} which describe how spacetime is sliced. The most commonly used gauge condition, the \textit{Gamma Driver} shift condition, imposes a restriction on the maximum time step size that is independent of the spatial resolution, as explained in Ref.~\cite{Schnetter_2010}. This restriction is still linked to the stability characteristics of the time integrator being used.

Given these considerations, we were led to ask whether it would be possible to design methods with fewer stages that retain the stability characteristics of Low Storage methods, leading to potentially larger ECFs. We have thus investigated methods based on a hybrid of the LM and RK methods, what is called a \textit{Multistep Runge-Kutta Method} (MSRK) in the literature. The idea is to use a combination of stage computations and previous time steps, thereby requiring fewer stage evaluations and obtaining a higher ECF. In this paper, we consider replacing one or two of the stages of the RK4 method with one or two previous RHS computations. The challenge is then to find the right coefficients for this new RK method that preserve the desired stability for NR problems we wish to study. We shall now describe the procedures we employed for obtaining and tuning the coefficients of such methods.

While there are many ways of introducing state reuse in an RK-like method, in this work, we shall restrict ourselves to the following class of 2-step method
\begin{align}
k_0 & = f\left(t_{n-1},\, y_{n-1}\right) \label{eq:update_two_step_start}\\
k_1 & = f\left(t_{n},\, y_{n}\right)\\
k_2 & = f\left[t_{n} + c_2 h,\, y_{n} + h \left(a_{20} k_0 + a_{21} k_1\right)\right]\\
k_3 & = f\left[t_{n} + c_3 h,\, y_{n} + h \left(a_{30} k_0 + a_{31} k_1 + a_{32} k_2\right)\right]\label{eq:update_two_step_last_k}\\
y_{n+1} & = y_n + h \left(b_0 k_0 + b_1 k_1 + b_2 k_2 + b_3 k_3 \right) \label{eq:update_two_step}
\end{align}
and the following class of 3-step method
\begin{align}
k_0 & = f\left(t_{n-2},\, y_{n-2}\right) \label{eq:update_three_step_start}\\
k_1 & = f\left(t_{n-1},\, y_{n-1}\right)\\
k_2 & = f\left(t_{n},\, y_{n}\right)\\
k_3 & = f\left[t_{n} + c_3 h,\, y_{n} + h \left(a_{30} k_0 + a_{31} k_1 + a_{32} k_2\right)\right]\label{eq:update_three_step_last_k}\\
y_{n+1} & = y_n + h \left(b_0 k_0 + b_1 k_1 + b_2 k_2 + b_3 k_3 \right) \label{eq:update_three_step}
.
\end{align}

We obtain the order conditions for $a_{ij}$, $b_i$ and $c_i$ by following the procedure outlined in Ref.~\cite{butcher_2006} for Runge-Kutta methods, based on using two Taylor series expansions around $h=0$. For the first, we use one of the update formulas, Eqs.~\eqref{eq:update_two_step} and \eqref{eq:update_three_step}; for the second, we use the function $y(t+h)$. Both expansions include terms of $\mathcal{O}(h^4)$. By collecting coefficients and matching the two series, we obtain a system of 8 equations and 11 variables for the 2-step method and a system of 8 equations and 8 variables for the 3-step method. To simplify these systems, we eliminate $b_0$ by applying the \textit{consistency condition}
\begin{equation}
  \sum_{i=0}^{3} b_i = 1.
  \label{eq:b_consistency}
\end{equation}
In addition, we apply the condition for $c_i$ given by
\begin{equation}
  c_i = \sum_{j=0}^{i-1}a_{ij}.
  \label{eq:c_condition}
\end{equation}

Solving the resulting equations for the remaining $a_{ij}$ and $b_i$, we obtain solutions that are parametrized by $c_2$ and $c_3$. For the 2-step method, the system has two 2-parameter solutions, which we denote with a $(1,2)$ superscript and are given by Eqs.~\eqref{eq:two_step_sol_start}--\eqref{eq:two_step_sol_end}. For the 3-step method, the system has one 1-parameter solution, given by Eqs.~\eqref{eq:three_step_sol_start}--\eqref{eq:three_step_sol_end}. We note that \rev{the TSRK method given by Butcher in {Ref.~\cite{butcher_2006}}, page 189, which we refer to as} Bu4-2, is a particular solution of Eqs.~\eqref{eq:two_step_sol_start}--\eqref{eq:two_step_sol_end}, obtained by choosing $c_2 = 1/2$ and $c_3 = 1$, whose coefficients we list in Table~\ref{tab:grand_coefficient_table}.

\subsection{Parameter tuning}\label{subsec:parameter_tuning}

Eqs.~\eqref{eq:two_step_sol_start}--\eqref{eq:three_step_sol_end} completely describe a member of the family of multistep methods once choices are made for the $c_2$ and $c_3$ coefficients. This choice is, in principle, arbitrary, but we have devised a scheme for finding the values of these constants that maximizes the value of the intersection of the ASR of the resulting method with the imaginary axis in a stability plot (which we refer to as the intercept value).

Our method of computing the intercept starts by obtaining the stability polynomials $\rho(\zeta;\;z)$ from Eqs.~\eqref{eq:update_two_step_start}--\eqref{eq:update_three_step_start} with the procedure described in Section~\ref{subsec:stability_of_ODE_solvers}. For our systems, it is then possible to find the set of roots $r_n(c_i;\;z)$ of $\rho(\zeta;\;z)$ analytically. By substituting Eqs.~\eqref{eq:two_step_sol_start}--\eqref{eq:three_step_sol_end} into $r_n(c_i;\;z)$, the inequality conditions for the ASR, Eq.~\eqref{eq:asr_def}, can also be determined analytically. Since we are interested in purely imaginary values that satisfy Eq.~\eqref{eq:asr_def}, we substitute $z \to B i$, where $i$ is the imaginary unit, and $B$ is an arbitrary real number. We then construct a bracketing scheme that takes $c_i$ and $B$ as inputs and returns $1$ if the condition is satisfied or $-1$ otherwise. This essentially turns the root condition for the ASR into a step function that transitions from $-1$ to $1$ precisely where the region crosses the real axis.

To find the $c_i$ that maximizes the intersection of the ASR with the imaginary axis, we must first choose a trial interval that the $c_i$ values can take. In our searches, we have chosen $c_i \in [-2,2]$. We then discretized these intervals with a total of $400$ points each and constructed lists of trial $c_i$. For example, in RK4-2 we have created a list of $(c_2, c_3)$ points with $400$ points in each dimension and values in the $[-2,2]$ interval, totaling $160,000$ trial values.

Once the trial $c_i$ values have been obtained, we compute the intercept for each $c_i$ using the routine described above. We then save these results in a list of $(c_i, \iota)$, where $\iota$ are the obtained intercept values. This repeated evaluation of the intercept function is trivially parallelizable, and doing so greatly speeds up the parameter search. Once all values are computed, we sort the resulting list by $\iota$ in descending order. The first element in this sorted list is therefore the largest ASR intercept obtained by the parameter search.

To ensure reasonable values, whenever a particular choice of $c_i$ would lead to $|a_{ij},b_i| > \delta$, where $\delta$ is an arbitrary bound, the set of coefficients $c_i$ is discarded from the search. In this paper, we have chosen $\delta = 4$, which was empirically determined to produce good results.

In Table~\ref{tab:grand_coefficient_table}, we present the numerical values of the coefficients that maximize the intercept value of the ASR obtained with the procedure described above. The first two columns of Tab.~\ref{tab:grand_coefficient_table} correspond to the coefficients for solutions $(1)$ and $(2)$ of the two-step method given in Eqs.~\eqref{eq:update_two_step_start}--\eqref{eq:update_two_step}. The third column corresponds to the coefficients for the three-step method of Eqs.~\eqref{eq:update_three_step_start}--\eqref{eq:update_three_step}. In the last column, we provide the coefficients for Bu4-2, for completeness.

\begin{figure}[ht]
  \centering
  \includegraphics[width=0.5\linewidth]{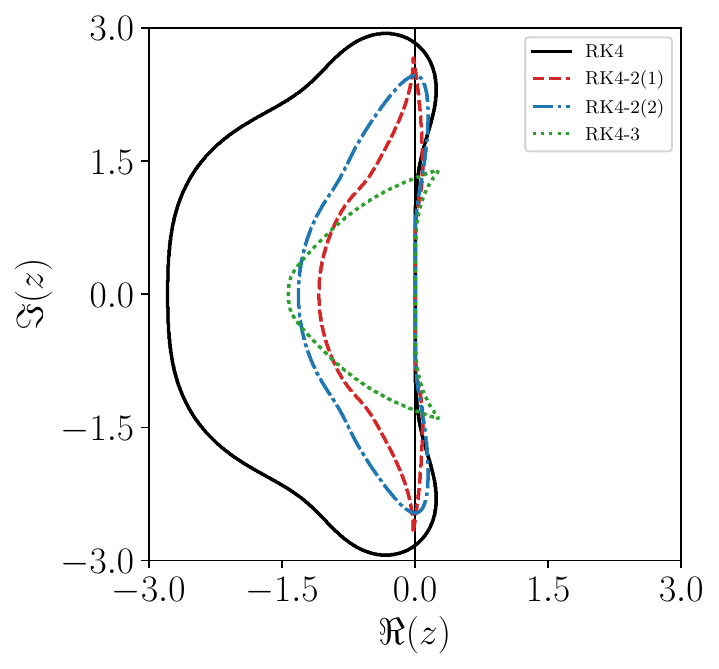}
  \caption{Absolute stability regions (ASRs) for \revd{RK4-2(1), RK4-2(2) and RK4-3 schemes given by the coefficients in Table~\ref{tab:grand_coefficient_table}. For comparison, we also plot the ASR of the standard RK4 method. The $x$ and $y$ axes represent, respectively, the real and imaginary parts of $z$, denoted by $\Re(z)$ and $\Im(z)$.
}}
  \label{fig:stability_compare}
\end{figure}

For RK4-2, the maximum ASR intercept values obtained were $2.53865$ for solution $(1)$ and $2.46201$ for solution $(2)$, while for RK4-3, the maximum intercept value was found to be $1.30711$. We have plotted the corresponding ASRs for RK4-2(1,2) and RK4-3 in \revd{Fig.~\ref{fig:stability_compare}. For reference and comparison purposes, we also plot the ASR of the RK4 method and note that the ASR intercept value for the standard RK4 method is given in {Ref.~\cite{peter_wave_eq}} as $\sqrt{8} \approx 2.8284$.}

\begin{table}
  \centering
  \caption{ASR intercept maximizing coefficients for solution (1) and (2) of the two-step method described in Eqs.~\eqref{eq:update_two_step_start}--\eqref{eq:update_two_step} \rev{(which we call RK4-2(1,2), respectively)} and~\eqref{eq:update_three_step_start}--\eqref{eq:update_three_step} (which we call RK4-3). The maximum ASR intercept values for RK4-2(1), RK4-2(2), and RK4-3 are respectively $2.53865$, $2.46201$, and $1.30711$.}
  \label{tab:grand_coefficient_table}
  \begin{tabular}{ccccc}
    \toprule
    Coefficient & RK4-2(1) & RK4-2(2) & RK4-3 & Bu4-2 \\
    \midrule
    $b_0$    & $-643/1536$      & $-191/882$          & $-85/1416$    & $0$    \\
    $b_1$    & $-4237/1092$     & $48241/59994$       & $131/408$     & $1/6$  \\
    $b_2$    & $38125/10752$    & $193750/4351347$    & $-29/24$      & $2/3$  \\
    $b_3$    & $4375/2496$      & $100000/271791$     & $15625/8024$  & $1/6$  \\
    $a_{20}$ & $-49/1250$       & $1309/15500$        & ---           & $-1/8$ \\
    $a_{21}$ & $399/1250$       & $-31999/15500$      & ---           & $5/8$  \\
    $a_{30}$ & $7033/960000$    & $-241289/5880000$   & $2511/62500$  & $1/2$  \\
    $a_{31}$ & $-217633/210000$ & $22846301/16170000$ & $-2268/15625$ & $-3/2$ \\
    $a_{32}$ & $5473/10752$     & $-936169/2587200$   & $29061/62500$ & $2$    \\
    $c_2$    & $7/25$           & $-99/50$            & ---           & $1/2$  \\
    $c_3$    & $-13/25$         & $101/100$           & $9/25$        & $1$    \\
  \bottomrule
\end{tabular}
\end{table}

\section{Numerical Tests}\label{sec:numerical_tests}

We have implemented RK4-2(1), RK4-2(2), and RK4-3 on the \texttt{Einstein Toolkit}\cite{roland_haas_2024_14193969} software infrastructure for numerical relativity and performed a series of standard and practical tests with the new time integrators. We've done this by extending the capabilities of \texttt{CarpetX}~\cite{schnetter2024carpetx}, the \texttt{AMReX}~\cite{zhang2019}--backed GPU-enabled \texttt{Cactus}~\cite{Goodale:2002a} driver.

Before we continue, it is important to discuss the practical matter of initializing and maintaining past RHS information in our MSRK methods. There are two situations where this comes into play. The first is at initialization, where no previous steps exist and therefore no past RHS information is available. The second is when regridding of an adaptive mesh takes place. When \texttt{CarpetX} changes the grid hierarchy, either because of tracking an object or refining the mesh in a certain region, the stored previous RHS information becomes invalid, since it corresponds to the older grid hierarchy. We remedy both situations by taking as many regular RK4 steps as necessary to fill the missing information. In the next sections, we will demonstrate that this strategy still produces consistent and convergent behavior.

Finally, we note that the tests presented in Secs.~\ref{sec:rst} and~\ref{sec:bbh} were executed on \texttt{Deep Bayou}\footnote{\url{https://www.hpc.lsu.edu/docs/guides.php?system=DeepBayou}}, a GPU cluster at Louisiana State University with 11 nodes, each containing two 24-core Intel Cascade Lake (Intel® Xeon® Gold 6248R Processor) CPUs and 2 NVIDIA V100S GPUs (with 32 GB of memory each). Each test fully utilized a single node of the machine and ran on the node's GPUs.

\subsection{Convergence}\label{sec:conv}

\begin{figure}[ht]
  \centering
  \includegraphics[width=\linewidth]{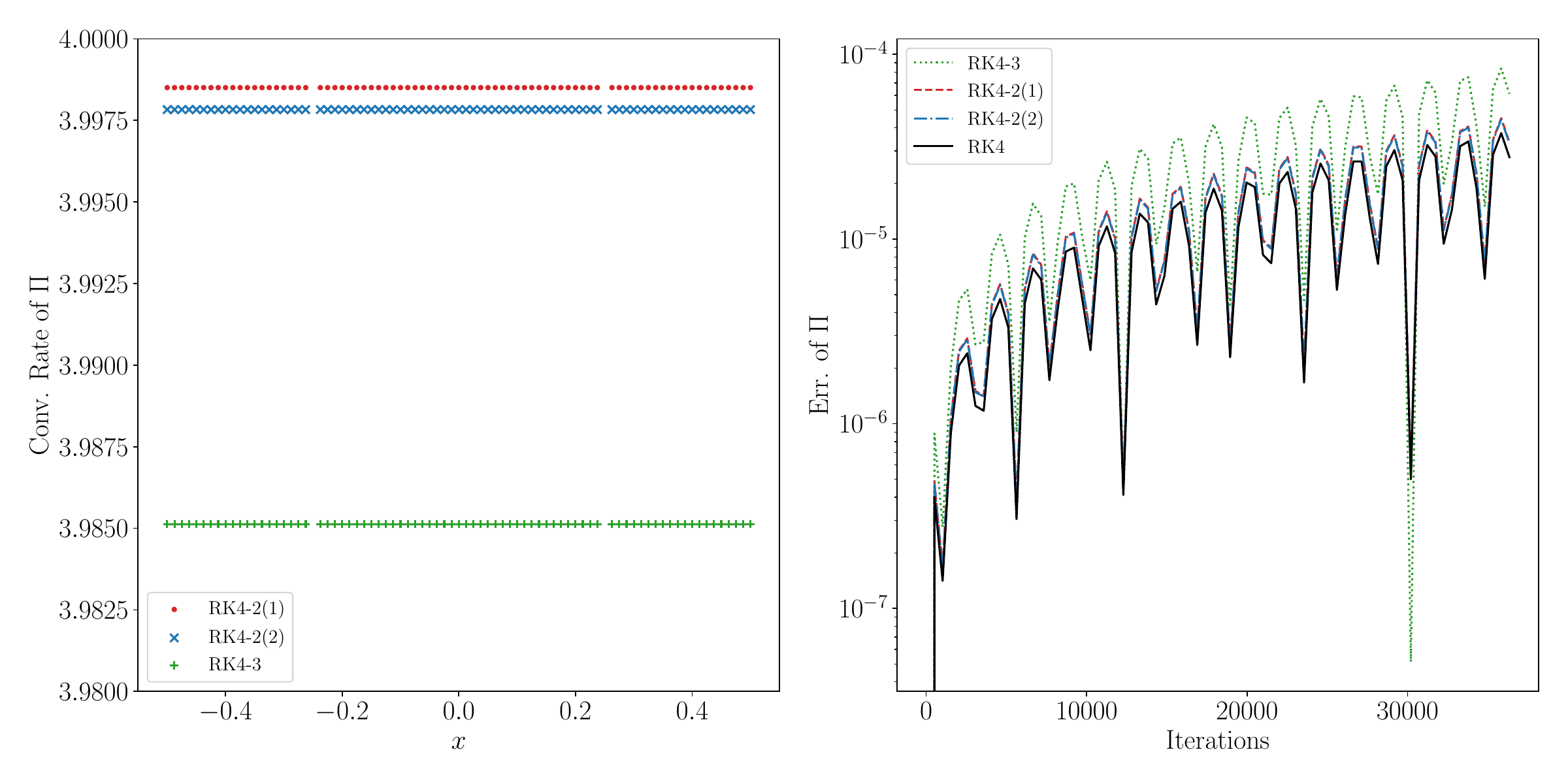}
  \caption{\revd{Left: Convergence tests for the evolution of a scalar wave equation using our new methods. Right: Evolution of the $\ell_\infty$ norm of the error in $\Pi$ versus the number of time steps taken (which was $36800$ or $100$ oscillations), computed with RK4 and our methods. The amplitude of the wave is $A=1$.}}
  \label{fig:conv_test}
\end{figure}

\revd{To demonstrate fourth-order convergence of our time integrators and the consistency of our startup procedures (the filling of missing past time levels by taking one or two RK4 steps at initialization), we evolve the standard scalar wave equation on a flat background for a field $\phi$ in first order form~\cite{Diener2007},

\begin{align}
    \partial_t \phi & = \Pi \\
    \partial_t \Pi & = \partial_i(\delta^{ij}d_j)\\
    \partial_t d_i & = \partial_i \Pi
\end{align}
where $\delta_{ij}$ is the Kronecker delta symbol.

We choose a standing wave profile for our initial data,
\begin{equation}
    \phi(t,x,y,z) = A \cos(2\pi\omega t)\cos(2\pi k_x x)\cos(2\pi k_x y)\cos(2\pi k_x z),
\end{equation}
where $A$ is the wave amplitude and $\omega = \sqrt{k_x^2 + k_y^2 + k_z^2}$.


We perform 3D evolutions with 4th-order accurate centered finite difference stencils and periodic boundary conditions. We choose domain extents $x,y,z \in [-0.5, 0.5]$ and set $A = k_{x,y,z} = 1$ for the initial data. With these choices, the standing wave completes a period of oscillation every $t = 1 / \omega = 1/ \sqrt{3}$. For all simulations, we choose a fixed CFL factor of $0.5$ and evolve the system until $4$ full oscillation periods have been completed. We repeat every simulation 3 times, with three resolution levels, using $80^3$, $160^3$ and $320^3$ grid cells respectively. We compute the convergence rate of $\Pi$ on 1D slices where $y=z=0$, with final iterations of  $368$, $736$ and $1472$ iterations, respectively. See Fig.~\ref{fig:conv_test}. Furthermore, we perform long-term evolutions of this system using our highest resolution level for $100$ oscillation periods ($36800$ iterations). We extract the $\ell_\infty$ norm of the error between the numeric and analytic solutions for $\Pi$ and plot it in the right panel of Fig.~\ref{fig:conv_test} for our methods and RK4. }

\subsection{Robust Stability Test}\label{sec:rst}

The Robust Stability Test (RST) was proposed in Refs.~\cite{rst_1,rst_2,rst_3,rst_4,rst_5}
and has been used as a pass/fail criteria for many NR evolution codes. The \revd{cited references vary significantly in the way the measurements are performed and the criteria for passing. For this work, we will primarily follow Ref.~\cite{rst_3}.}

The test measures both the well-posedness of a continuum problem and the stability of the numerical algorithms being used to discretize and solve it.
Given some appropriate discrete norm $\lVert v_n \rVert$ at time step $t_n$, the RST is passed if there are constant values of $A$ and $K$ such that
\begin{equation}
  \frac{\lVert v_n \rVert}{\lVert v_0 \rVert} \leq A e^{K t_n}
  \label{eq:rst_def}
\end{equation}
is satisfied independently of the choice of initial data for all possible spatial grid spacings $h$. In practice, it is impossible to inspect results for all possible spatial resolutions, and thus one looks for trends in the data as the resolution increases. If the rate of growth in the norm data increases with resolution, the test is said to have \textit{failed}, otherwise it is said to have \textit{passed}. In our case, we utilize the RST to both assert the correct implementation of our multistep methods and to gauge the highest CFL factor that can be attained for a passing (stable) evolution.

We perform our evolutions in a ``long channel,'' following the example of the RST papers above. We create a grid whose extent is \revd{$x,y,z \in [-0.5, 0.5]$ and contains $n_{x} \rho$ grid points in the $x$ direction, where $\rho = 1,2,4$, and $n_{yz} = 10$ grid points in the remaining direction. We choose this value because it is slightly in excess of the minimum number of grid points required for a fourth-order evolution.}

Our physical system of interest is
Einstein's field equations of general relativity in BSSN (Baumgarte-Shapiro-Shibata-Nakamura) form (see Ref.~\cite{bssn_1, bssn_2} for details). To evolve the BSSN system, we employ \texttt{CanudaX\_LeanBSSNMoL}~\cite{CanudaX_repo}, a \texttt{CarpetX} port of \texttt{Canuda}'s \texttt{LeanBSSNMoL}~\cite{canuda} code. We initialize the system with the Minkowski flat spacetime solution (i.e. empty space) and \revd{add a small amount of pseudo-random noise $\epsilon$ to initial data}. The noise amplitude is chosen to be in the range given by
\begin{equation}
  \epsilon \in \left[-10^{-10} / \rho^2, 10^{-10}/\rho^2\right],
  \label{eq:noise_range}
\end{equation}
which ensures that the initial error introduced by the noise will, on average, be the same for each resolution factor $\rho$. Following Ref~\cite{rst_3}, we evolve the system for 1 crossing time, which in this case corresponds to $t=1$ (simulation time) and save data at every time step. The test is repeated with different CFL factors until a passing result is obtained.

A few remarks on the choice of pseudo-random noise generation are in order. For all our test runs, we employ the standard \texttt{C++ random} library using the \texttt{std::mt19937} random engine, with integer seed fixed at $100$. All random values are obtained from a \texttt{std::uniform\_real\_distribution} in the range given in Eq.~\eqref{eq:noise_range}. All \texttt{random} objects are initialized globally and only once during program startup. This ensures the predictability and reproducibility of all the pseudo-random numbers employed in our tests. Note, however, that by increasing the resolution factor $\rho$, more random numbers will be produced by the generator. This in turn means that two resolutions can't ever be initialized with exactly the same numbers, and thus, shall never have the same initial values for $\lVert\mathcal{H}\rVert_2$. In fact, by keeping $\rho$ fixed and changing the random seeds, one can observe variations in $\lVert\mathcal{H}\rVert_2$, which gets shifted along the $y$ axis by constant values. 

\begin{figure}[ht]
  \centering
  \includegraphics[width=\linewidth]{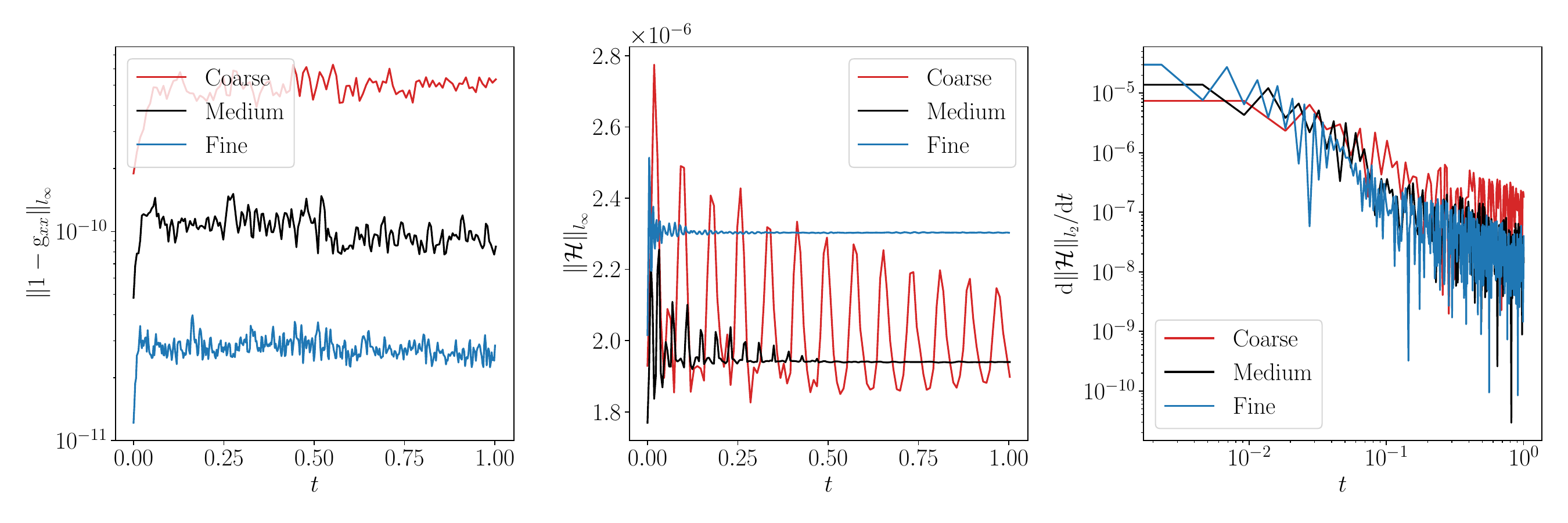}
  \caption{Example RST results for a passing CFL of $0.46$ using RK4-2(1).}
  \label{fig:rst_HRK423_pass_fail}
\end{figure}

To eliminate this constant offset, we elect to analyze the time derivative of $\lVert\mathcal{H}\rVert_{\ell_2}$.
This solves the offset problem and does not hinder the RST in any way, since all we are interested in is analyzing the growth rate of the norm in time, and thus, its derivative. In other words, a convergent code will obey
\begin{equation}
  \frac{\mathrm{d} \lVert\mathcal{H}\rVert_{\ell_2}}{\mathrm{d}t} \approx 0,
  \label{eq:converge}
\end{equation}
while a divergent one will grow without bound. \rev{Additionally, we also study the behavior of ``standard'' RST norms, i.e., the error norm $\lVert \delta_{ij} - \mathrm{g}_{ij}\rVert_{\ell_\infty}$ and $\lVert\mathcal{H}\rVert_{\ell_{\infty}}$. For illustrative purposes, the result of the passing RST for the RK4-2(1) with CFL factor $0.46$ is presented in Fig.~\ref{fig:rst_HRK423_pass_fail}. Note that in a passing RST, the time derivative of the Hamiltonian constraint norm converges to zero as resolution increases. We summarize in Table~\ref{tab:result_summary} (under the RST column) the passing CFL / ECF value pairs for each of our tested methods.}

\subsection{Binary Black Holes}\label{sec:bbh}

In this test, we have evolved a Binary Black Hole (BBH) system in fully nonlinear numerical relativity using RK4-2(1), RK4-2(2), and RK4-3 while comparing several evolution metrics with a reference RK4 evolution of the same system. 

Binary black hole initial data was provided by \texttt{TwoPuncturesX}, a \texttt{CarpetX} port of the \texttt{TwoPunctures}~\cite{two_punctures} spectral initial data solver in the standard \texttt{qc0} configuration, i.e. a collision of two black holes of equal mass $0.453$, equal momenta $\pm 0.3331917498$, and located at grid coordinates $(x = \pm 1.168642873,\, y = 0,\, z=0)$. For our system of units, \revd{we use $c = G = 1$ and the mass is given in solar masses.} The movement of the punctures was tracked using the \texttt{PunctureTracker} thorn. 

The specific software modules that enable us to extract this wave are the \texttt{WeylScal4}~\cite{WeylScal4_repo, PhysRevD.65.044001} and \texttt{Multipole}~\cite{Multipole_repo, Multipole_docs} thorns.

Boundary conditions were provided by \texttt{NewRadX}, a \texttt{CarpetX} port of \texttt{NewRad}~\cite{new_rad}. We have also employed 5th order Kreiss-Oliger artificial dissipation, built into \texttt{CanudaX\_LeanBSSNMoL}. 

All simulations were run up to $t=500$ (simulation time, approximately $2.5$ milliseconds), within a cubical grid structure with $7$ mesh refinement levels arranged in a ``box in box'' style, with the inner boxes having the highest resolutions and following the movement of the punctures throughout the evolution. The outermost (coarsest) level extended from $-120$ to $120$ (i.e. about $350$ km across), and had $120$ cells on all 3 spatial directions. 

The grid structure moves throughout the simulation in such a way that the centers of the finer levels approximately coincide with the centers of the punctures. Here, we remind readers of the discussion in the beginning of Sec.~\ref{sec:numerical_tests}: whenever the grid structure changes, it is necessary to refill previous RHSs by taking a number of regular RK4 steps.

\revd{Using a fixed CFL factor of $0.45$}, we performed simulations for RK4, RK4-2(1), and RK4-2(2), allowing us to compare the performance of these methods directly. The CFL $0.3$ was selected for RK4-3, however, because it could not successfully run with a greater value. This value is larger than the maximum CFL determined via RST in Sec.~\ref{sec:rst} and is made possible thanks to the artificial dissipation added to the system. It is important to note that we did not use subcycling in time for any of the runs presented in this paper. This means that \texttt{CarpetX} selects a single time step for all levels which honors the requested CFL condition on the finest level.

\begin{figure}
    \centering
    \includegraphics[width=\linewidth]{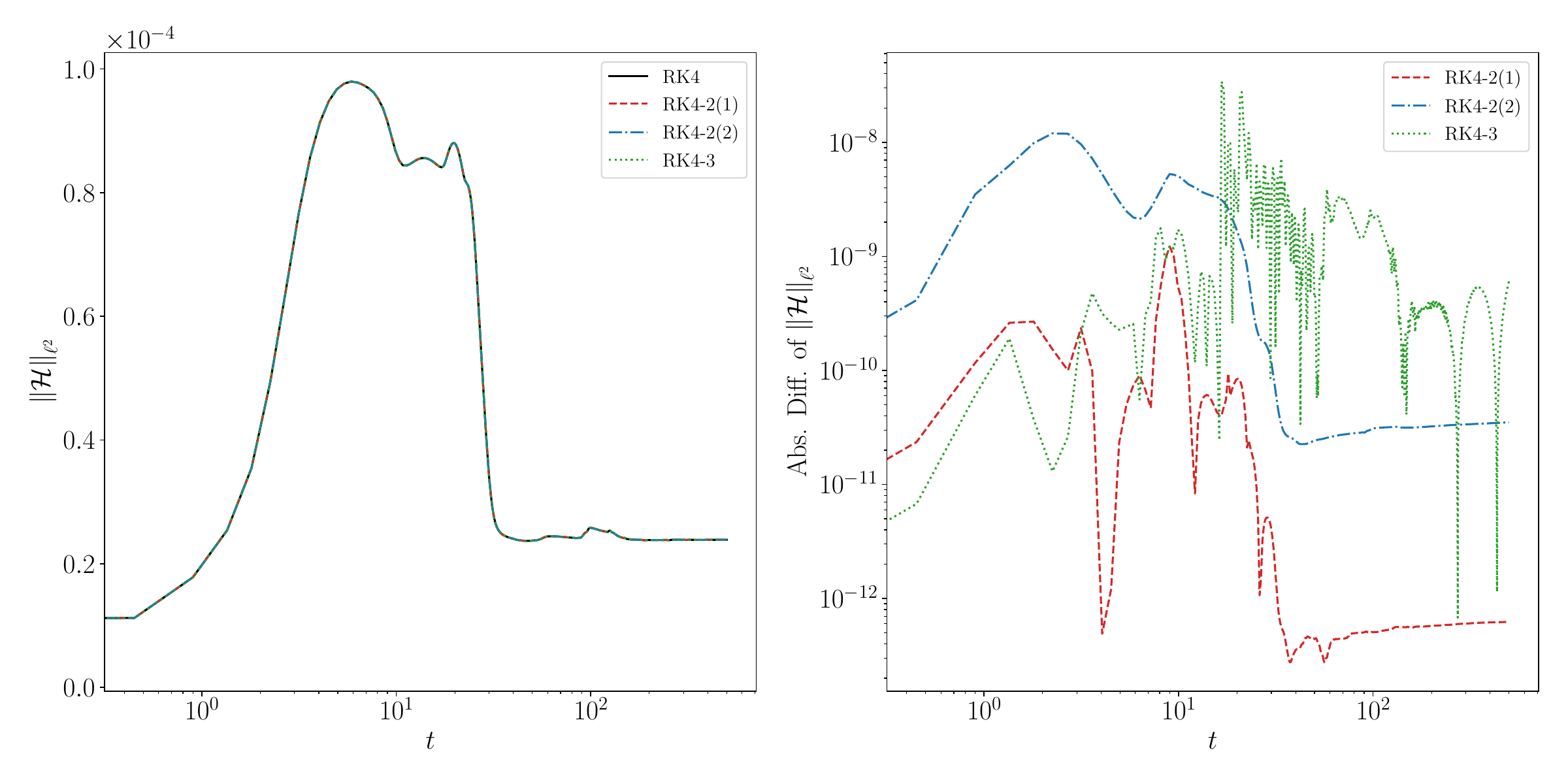}
    \caption{\revd{Log-log plots of the time evolution of the $\ell^2$ norm of the Hamiltonian constraint (left) and a log-log plot of the relative differences of that quantity obtained with our methods to that obtained with RK4 (right)}}
    \label{fig:bbh_hc_evol_and_diff}
\end{figure}

The first criteria used for gauging the quality and correctness of our methods was to measure the evolution of the of the $\ell^2$ norm of the Hamiltonian constraint $\lVert\mathcal{H}\rVert_{\ell^2}$ (which, mathematically, should always be zero) obtained with RK4 and our methods. \revd{The left panel of Fig.~\ref{fig:bbh_hc_evol_and_diff}, shows the evolution of $\lVert\mathcal{H}\rVert_{\ell^2}$ for our methods and RK4, while the right panel shows the relative differences of the same quantity between our methods and RK4 on a log-log plot. Even though all methods present good agreement with the results found by RK4, RK4-2(1)'s overall performance in this test is the best} of all methods.

\revd{Next, we perform a similar comparison with the extracted $r=90,\, l=2,\, m=2$ multipolar component of the $r\times\Psi_4$ Weyl scalar. The left panel of Fig.~\ref{fig:bbh_hc_evol_and_diff} displays the extracted mode for our methods and RK4, while the right panel shows the relative difference of the signal obtained with our methods with respect to that obtained using RK4 on a log scale in the $y$ axis. Once again, RK4-2(1) reveals itself to be the best performing of all the methods.

\begin{figure}
    \centering
    \includegraphics[width=\linewidth]{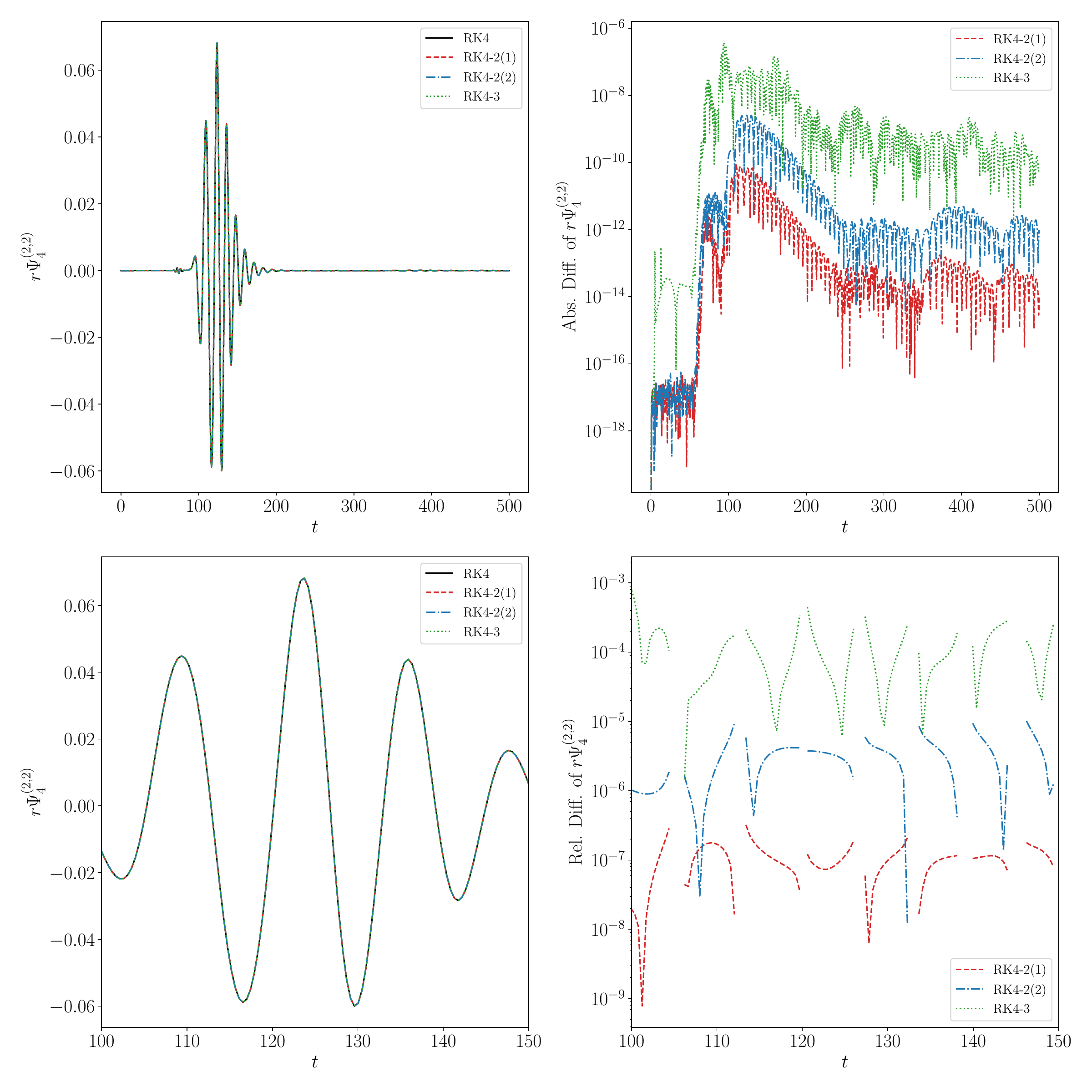}
    \caption{$r \Psi_4(2,2)$ at $r=90$ evolution obtained with our methods and RK4 (left) and relative differences differences of the same signal obtained with our methods with that obtained with R4K on a log scale in the $y$ axis}
    \label{fig:bbh_results}
\end{figure}
}

We have also conducted performance measurements of simulations using both RK4 and RK4-2(1) to determine the speedup achieved when using our MSRK method. Toward that end, we evolved the code for 64 time steps and computed the average cumulative time of various sections of our code. The number of time steps was chosen such that the performance measurements were untainted by factors such as IO and regridding. Each section was measured by a corresponding \texttt{Cactus} timer object, which is a scoped and MPI synchronized timer.

By comparing the timers between the two methods, we obtain a $30\%$ speedup when using RK4-2(1) relative to RK4. Even though this is a large speedup, it is still $3\%$ smaller than the theoretical maximum achievable speedup of $33\%$. To clarify, since RK4 requires $4$ RHS evaluations and RK4-2 requires $3$ RHS evaluations, one might expect a speedup of $1/0.75=1.33$ or $33\%$. However, because the overhead resulting from copying, linear combinations, etc. accounts for $8\%$ of the time spent in RHS evaluations, we get a speedup of $(1+.08)/(.75+.08)=1.3$ or $30\%$ \rev{with respect to walltime}.

\subsection{General Relativistic Magneto-hydrodynamics}\label{sec:grmhd}

In this section, we have used our methods to perform various test evolutions in the context of General Relativistic Magnetohydrodynamics (GRMHD) using \texttt{AsterX}~\cite{Kalinani_2025}, a new GPU enabled GRMHD code for \texttt{CarpetX}.

The RHS of the evolution equations in GRMHD applications is more complex than in the vacuum evolution of BBHs because these equations are solved using Finite Volume schemes, which evolve \textit{conserved} variables in time. These conserved variables, however, are not trivially correlated with the physically relevant \textit{primitive} quantities. Therefore, GRMHD codes have to convert from conserved variables to primitive variables at each RHS evaluation. This involves an iterative root-finding step that takes place at every grid point.

The first test we performed was a 1D relativistic shock-tube. This test is a standard correctness test from the test suite introduced in Ref.~\cite{Balsara_2001}, known as the \textit{Balsara} test suite. For illustrative purposes, we chose to display only the fluid density, $\rho$ for the \texttt{Balsara 3} test. This particular test was chosen because it is the most demanding of the tests, thanks to the sharp jump in initial fluid pressure.

We performed the test on a cube with extents of $[-0.5, 0.5]$. We used $400$ cells across the $x$ direction of the grid and $2$ cells across the $y$ and $z$ directions. We used an ideal gas equation of state with \texttt{minmod} reconstruction and fluxes computed by the \texttt{HLLE} approximate Riemann solver. We evolved the system until $t=0.4$ \revd{with a CFL factor of $0.25$ and compared the results with the problem's analytical solutions, obtained with the exact Riemman problem solver described in Ref.~\cite{GIACOMAZZO_REZZOLLA_2006}. The tests were run with RK4 and our methods under the same settings.}

\begin{figure}[ht]
  \centering
  \includegraphics[width=\linewidth]{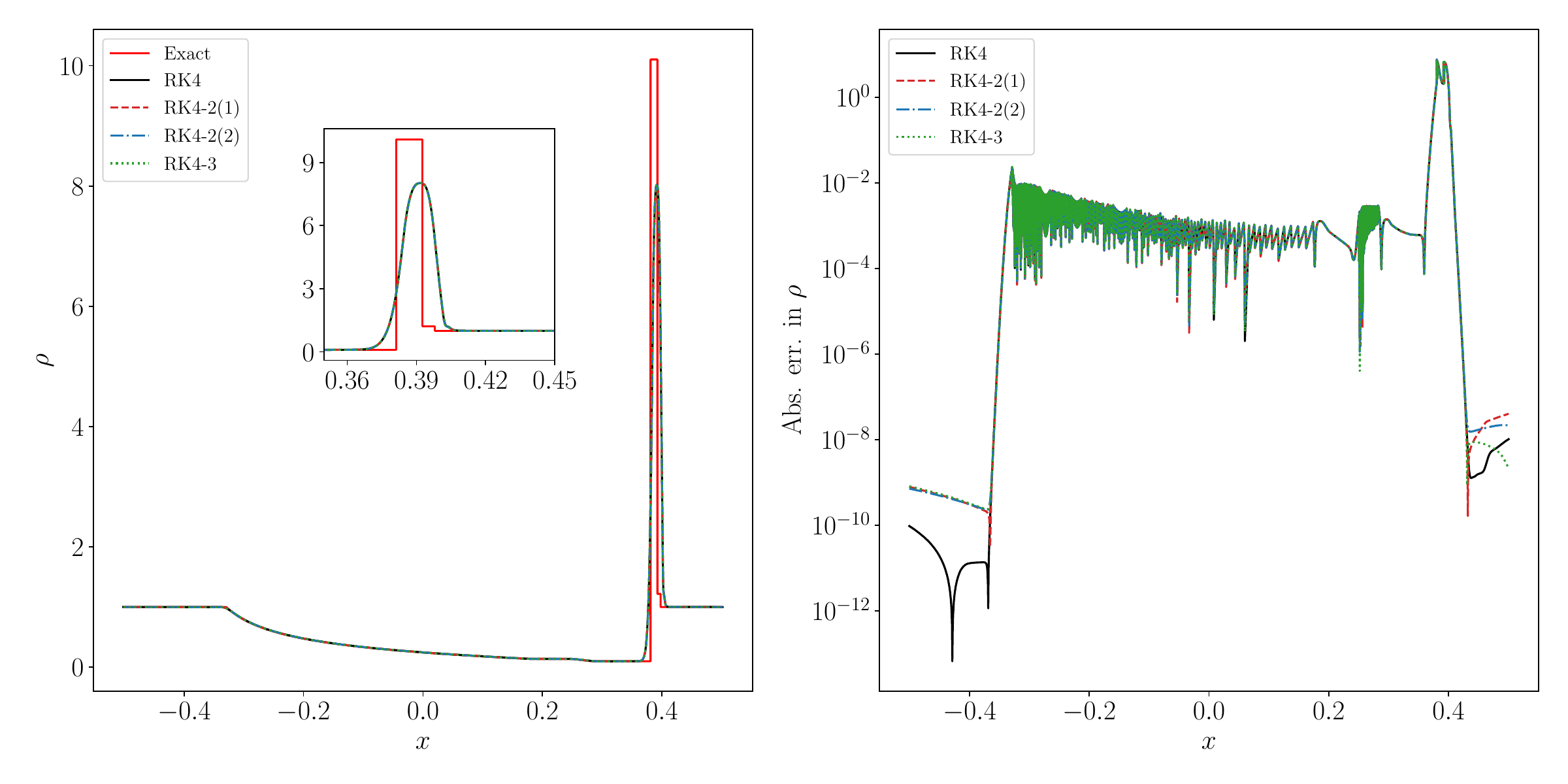}
  \caption{Left: \texttt{Balsara 3} tests results \revd{for the fluid density $\rho$ using RK4, comparing our methods and the test's exact solution. Right: The relative differences between the solutions obtained using our methods and RK4 to the exact solution.}}
  \label{fig:balsara3}
\end{figure}

In Fig.~\ref{fig:balsara3}, we show on the left panel the \texttt{Balsara 3} test results obtained with \revd{our methods and RK4 as well as the exact solution. On the right panel, we show the relative error of the solutions obtained with all methods against that of the analytic solution using a log scale on the $y$ axis. The plot reveals good agreement between the results obtained with our new methods and those obtained with RK4. In particular, we found that RK4-3 yielded the smallest maximum relative difference, across all methods.}

In addition to the \texttt{Balsara 3} test, we performed a Kelvin-Helmholtz Instability (KHI) evolution using our proposed methods. The KHI is a phenomenon that occurs when a velocity difference is present between two fluid layers (or in the interface between two different fluids), which causes the fluid to transition into a turbulent behavior. 

\revd{The KHI instability test is more complex than the \texttt{Balsara 3} test. Our setup follows that of Ref.~\cite{Kalinani_2025} closely. We perform the test in 2D, and use the AMR capabilities of \texttt{CarpetX} using $4$ refinement levels with $64^2$ grid cells on the coarsest level. Once again we employ the $[-0.5, 0.5]$ coordinate domain in the $x$ and $y$, directions, with periodic boundary conditions across the $x$ direction. We chose a fixed CFL factor of $0.25$ and evolved the system up to $t=1.5$. We initialize a central ``band'' of fluid in the $|y| < 0.25$ region with density $\rho = 2.0$ and velocity $v_x = 0.5$ across the $x$ direction. The surrounding fluid is initialized with density $\rho = 1.0$ and velocity $v_x = - 0.5$. This setup provides a discontinuous interface between the central ``band'' and its surroundings. To induce the instability, we introduce a slight perturbation across the velocity of the fluid in the $y$ direction.

It is important to note that, unlike Ref.~\cite{Kalinani_2025}, we have employed the Lax-Friedrichs flux (LxF) with minmod recostruction for the solution of the KHI. This was necessary in order to successfully run the low storage methods. Using HLLE with the low storage method simulations, an instability arises in the density which causes the code to terminate. We believe that the extra dissipative nature of LxF is needed to help these schemes achieve stability. Finally, we also note that we have observed that our methods, as well as standard RK4, encountered no difficulty evolving this system with HLLE obtained fluxes and the use of LxF seemed to produce no difference in the maximum achievable CFL of our methods (or of RK4) for this problem.

In Fig.~\ref{fig:khi_full}, we plot the final iteration of the evolution, demonstrating the fully developed instability for RK4 and our methods.

\begin{figure}[ht]
  \centering
  \includegraphics[width=0.7\linewidth]{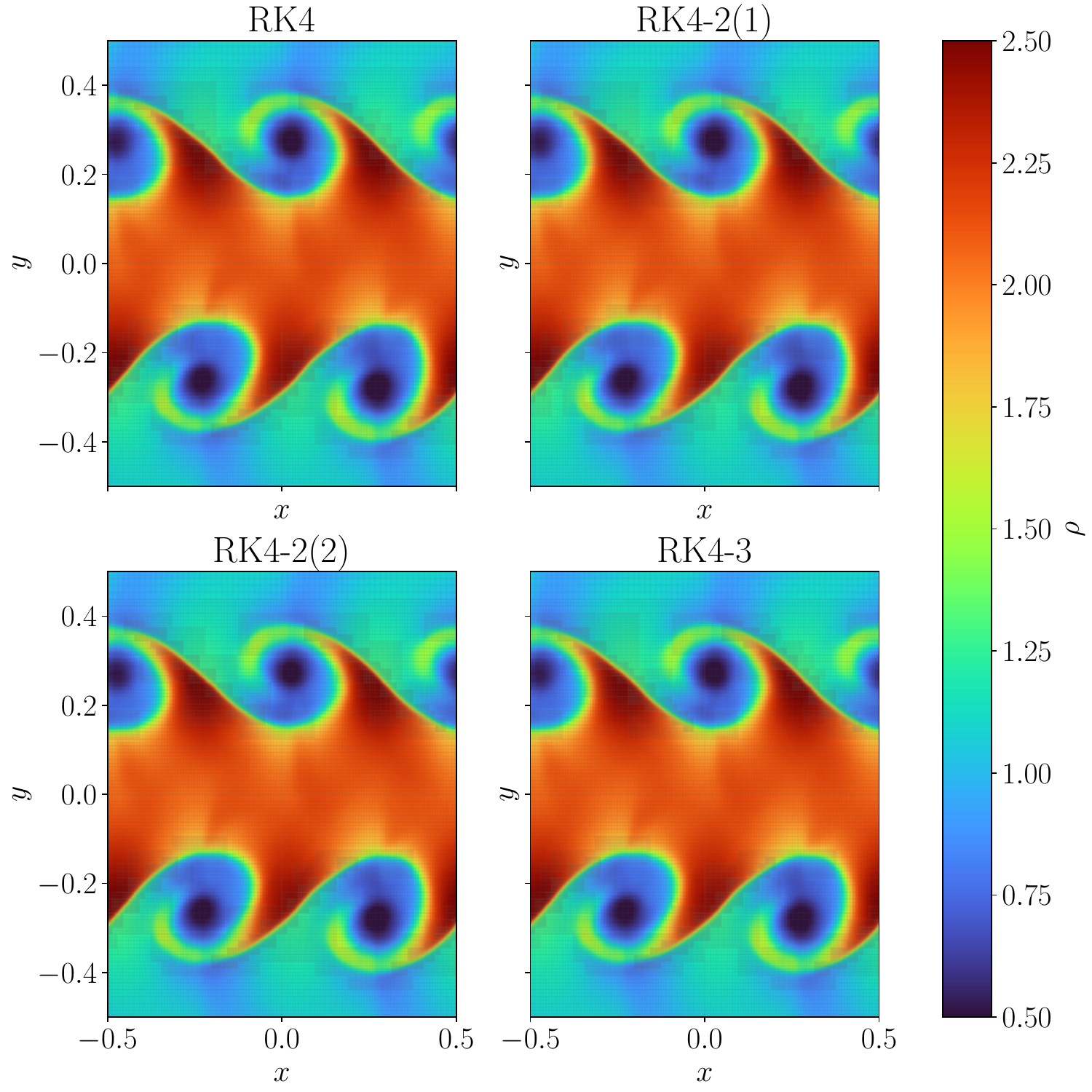}
  \caption{\revd{Fully developed KHI evolution fro RK4 and our new methods.}}
  \label{fig:khi_full}
\end{figure}
}

\subsection{Maximum CFL analysis}\label{sec:max_cfl}

\revd{We have also sought to determine the maximum CFL and ECF values that could be used in all of the tests described so far. This analysis can help reveal which method is the most efficient when solving a particular problem. For completeness, we have included in this analysis two Low Storage methods: RK4(3)6[2S] from Ref.~\cite{Ketcheson_2010} and RK4(9[3S*] from Ref.~\cite{Ketcheson_2013} (in this reference, this method is referred to as ERK(9,4)). These specific methods were selected for being the ones with intercept values larger than those of RK4 and thus having the greatest potential for having ECFs that provide a speedup (with respect to RK4).

\begin{table}[th]
  \centering
  \caption{Maximum CF / ECF values for successfully evolving a system with a given method.}
  \begin{tabular}{cccccc}
    \toprule
    Method          & WEQ              & RST              & Balsara3         & KHI              & BBH              \\
    \midrule
    RK4-2(1)        & $1.14$ / $0.380$ & $0.46$ / $0.153$ & $0.37$ / $0.123$ & $0.39$ / $0.130$ & $0.47$ / $0.157$ \\
    RK4-2(2)        & $1.05$ / $0.380$ & $0.44$ / $0.147$ & $0.45$ / $0.153$ & $0.41$ / $0.137$ & $0.59$ / $0.197$ \\
    RK4-3           & $0.57$ / $0.285$ & $0.23$ / $0.115$ & $0.42$ / $0.210$ & $0.45$ / $0.225$ & $0.30$ / $0.150$ \\
    RK4             & $1.22$ / $0.305$ & $0.51$ / $0.127$ & $0.94$ / $0.235$ & $0.86$ / $0.215$ & $0.70$ / $0.175$ \\
    RK4(3)6{[}2S{]} & $1.89$ / $0.315$ & $0.79$ / $0.132$ & $0.60$ / $0.100$ & $1.00$ / $0.167$ & $1.00$ / $0.167$ \\
    RK4()9{[}3S*{]} & $2.17$ / $0.241$ & $0.91$ / $0.101$ & $0.85$ / $0.094$ & $2.38$ / $0.264$ & $1.55$ / $0.172$ \\
    \bottomrule
  \end{tabular}
  \label{tab:result_summary}
\end{table}

\subsubsection{Wave Equation (WEQ)}

For this test, an analytic solution is known. Given a choice of method and CFL factor, we initialized the system with standing wave initial data and evolved it for 3 crossing times, using the same grid setup described in Sec.~\ref{sec:conv}. Once the evolution was completed, we examined the average error of the solution, obtained by comparison with the known exact solution for this test. If this value was below our tolerance threshold (which in this case, we chose to be $1.0\times10^{-2}$), we consider the test to have passed. Otherwise, the test is considered to have failed. We explore the parameter space of CFL factors by employing an iterative binary search. We choose bounds for the parameter space using values that are guaranteed to fail for the maximum bound and values that are guaranteed to work for the minimum bound. Specifically, we have chosen these values to be $4.0$ and $0.1$, respectively. At every iteration of the search algorithm, we perform the test with CFL given by $\text{min} + (\text{max} - \text{min})/2$. If the test passes, we update the minimum bound to be the current passing CFL. If not, we reduce the maximum bound to be the current failing CFL. This test is performed for 20 iterations, which we found to be enough to provide converging CFL values. 

The results of this parameter search are presented in the ``WEQ'' column of Table~\ref{tab:result_summary}, where we list CFL / ECF.

\subsubsection{Robust Stability Test (RST)}

Because this test already has a clearly defined pass/fail criteria (See Appendix A.1 of Ref.~\cite{rst_3}), and is a computationally cheap, 1D test, maximum CFL / ECF pairs can be obtained directly by manually running the test code with multiple different CFL values and determining the maximum.

The results of test are presented in the ``RST'' column of Table~\ref{tab:result_summary}, where we once again, list CFL / ECF pairs.

\subsubsection{Balsara 3}

For this test, since an analytic solution is known, we employed the same parameter search procedure we used on the WEQ test. In this test, however, we chose the tolerance threshold differently. First, we evolved the system under the ``standard'' conditions described in Sec.~\ref{sec:grmhd} using RK4 and computed the average of the absolute error norm to be approximately $0.20$. Then, we performed the bisection search algorithm using a tolerance of $1.1 * 0.20$ in the average error norm. The CFL / ECF pairs of results for this test are presented in the fourth column of Tab.~\ref{tab:result_summary}

\subsection{Kelvin-Helmholtz instability (KHI)}

Given that no analytic solution for the KHI exists, we created a baseline evolution of the system using RK4 as a time integrator and fixed CFL factor of $0.25$. From this, we established baseline minimum and maximum density values achieved by the system at $t=1.5$. With these in hand, we have explored different CFL factors and considered results acceptable if they produced minimum and maximum densities that were within $1\%$ of the baseline values. We list the maximum passing CFLs, according to the criteria described, in the fifth column of Table~\ref{tab:result_summary}. To help illustrate that this procedure still results in sensible KHI evolutions, we show in Fig.~\ref{fig:khi_max} an image of the evolution obtained with RK4-3 and CFL according to Tab.~\ref{tab:result_summary}. Evolutions with the other methods and CFLs reported in the table look indistinguishable from that of Fig.~\ref{fig:khi_max}, thus we choose not to show them here for the sake of brevity.

\begin{figure}[ht]
  \centering
  \includegraphics[width=0.5\linewidth]{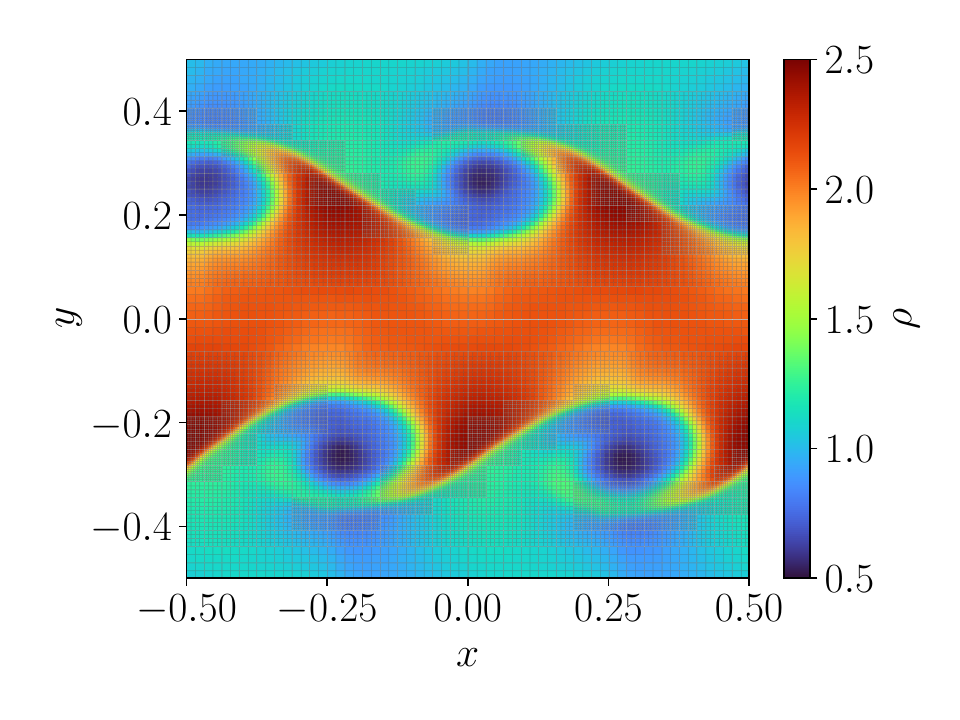}
  \caption{KHI evolution using RK4-3 and the maximum CFL value reported in Tab.~\ref{tab:result_summary}.}
  \label{fig:khi_max}
\end{figure}

\subsection{Binary black holes (BBH)}

Once again, no analytic solutions for this problem exist, and we again compared with the well-established standard RK4 time integrator result. For this analysis, we sought to determine which maximum CFLs could be used for a given time integrator and have it still produce ``sensible'' waveforms. To make this criterion more formal, we have repeated the test detailed in Sec.~\ref{sec:bbh} with different CFLs and methods. For each run, we computed the average of the squared difference between the extracted $l=m=2$ mode at $r=90$ of the $\Psi_4$ Weyl scalar obtained with RK4 and a given method. We consider the test passed if this norm is below $10^{-15}$, or failed otherwise. We found that for all reported values in Tab.~\ref{tab:result_summary}, all waveforms look visually indistinguishable.
}

\section{Dense Output Coefficients}\label{sec:dense_output}

The CFL condition imposes an upper limit on the time step size. It depends upon the problem, the time integrator, and the spatial discretization scheme being used. As a consequence, when using adaptive mesh refinement, each refinement level has different grid spacings and thus different time step restrictions that need to be satisfied. There are two approaches to solving this problem. The simplest but least computationally efficient way is \texttt{uniform stepping}. For this method, we choose a global time step that honors the CFL condition restriction imposed by the finest grid.

The second approach is called \texttt{subcycling in time}. This approach allows each level to set its own Courant factor, allowing coarser grids to take fewer steps than finer grids. A complexity of this approach is that it requires data to be filled in on the boundaries of fine grids. 

One solution for this problem, known as \textit{buffer zones}~\cite{Schnetter_carpet} or \textit{tapering}~\cite{Lehner_2006} was implemented in \texttt{Carpet}, Cactus's fixed mesh refinement framework. Broadly speaking, buffer zones extend a mesh refinement region by four times the width of its ghost zones and, during each step of RK4, progressively discarding outer zones as each RHS is evaluated. This solution is expensive, both for memory and compute cycles. In principle, if our RK4-2 or RK4-3 methods were to be used in this scenario, they would greatly reduce the number of such buffer zones. Conversely, the low storage methods would increase it.

For \texttt{CarpetX}, we use an approach called the Berger-Oliger without order reduction (BOR) algorithm. It was first described in Refs.~\cite{Mongwane2015,mccorquodale2011high}, and details of our implementation can be found in Ref.~\cite{carpetx_subcycling}. It involves creating a polynomial that interpolates the values of $y_n$ within a time step. This technique is known as \textit{dense output}, and it allows us to fill in inter-refinement level ghost points for the stages of the RK methods on the refinement levels. While dense output polynomials for Runge-Kutta have been known in the literature (where they are used for enhancing visualization, or stopping conditions for ODE solvers, See Ref.~\cite{shampine2015} for details), their use in adaptive mesh refinement contexts is more recent.

Although \texttt{CarpetX} does not yet have the necessary infrastructure to use our new method with subcycling, we can nevertheless provide the dense output polynomial our method would require. We note that this formula could be of use to any code which needs to interpolate the $y_n$ solutions within a time step.

We will now present our approach for computing dense output coefficients for the methods developed in Sec.\ref{sec:hybrid_method} following Refs.~\cite{horn1983, shampine2015}.

We seek to find \revd{the functions} $d_i$ such that
\begin{equation}
  \frac{\mathrm{d}y(t + \theta h)}{\mathrm{d}t} \approx \sum_{i=0}^{3} d_i(\theta) k_i
  \label{eq:dense_output_approx}
\end{equation}
for the $k_i$ given by either Eqs.~\eqref{eq:update_two_step_start}-\eqref{eq:update_two_step_last_k} for a two-step method or Eqs.~\eqref{eq:update_three_step_start}-\eqref{eq:update_three_step_last_k} for a three-step method.

To do this, we first recognize that
\begin{equation}
  \frac{\mathrm{d}y(t + \theta h)}{\mathrm{d}t} = f(y(t+\theta h))
\end{equation}
and expand both $f(y(t+\theta h))$ and the RHS of Eq.~\eqref{eq:dense_output_approx} around $h=0$ to third order. We match the coefficients of both expansions and obtain 4 equations with 4 unknown $d_i$ values. We substitute the set of coefficients for one of the multistep methods given in Tab.~\ref{tab:grand_coefficient_table} in the resulting equations and solve them for $d_i$, which depend on $\theta$. To find the update formula, we substitute $d_i(\theta)$ in Eq.~\eqref{eq:dense_output_approx} and integrate it with respect to $\theta$ to find an expression of the form
\begin{equation}
  y(t + \theta h) = y(t) + h \sum_{i=0}^{3}e_i(\theta) k_i
\end{equation}
where $e_i(\theta)$ are given explicitly in Eqs.\eqref{eq:two_step_dense_output_coeff_start}-\eqref{eq:two_step_dense_output_coeff_end} for RK4-2(1,2) and in Eqs.~\eqref{eq:three_step_dense_output_coeff_start}-\eqref{eq:three_step_dense_output_coeff_end} for RK4-3.

By applying this procedure, we obtain the dense output coefficients $e_i(\theta)$ for RK4-2(1)
\begin{align}
  e_0^{(1)}(\theta) &= -\nicefrac{643 \theta}{1536} \label{eq:two_step_dense_output_coeff_start}\\
  e_1^{(1)}(\theta) &= -\nicefrac{\theta (837+100 \theta (9+25 \theta))}{1092} \\
  e_2^{(1)}(\theta) &= \nicefrac{5 \theta (1929+64 \theta (39+50 \theta))}{10752} \\
  e_3^{(1)}(\theta) &= \nicefrac{5 \theta (643+8 \theta (-21+50 \theta))}{2496},
\end{align}
the dense coefficients $e_i(\theta)$ for RK4-2(2),
\begin{align}
  e_0^{(2)}(\theta) &= \nicefrac{\theta^2 (-291+100 \theta)}{882} \\
  e_1^{(2)}(\theta) &= \theta + \nicefrac{(4947-16700 \theta) \theta^2}{59994} \\
  e_2^{(2)}(\theta) &= \nicefrac{38750 \theta^2 (3+2 \theta)}{4351347} \\
  e_3^{(2)}(\theta) &= \nicefrac{20000 \theta^2 (3+2 \theta)}{271791},\label{eq:two_step_dense_output_coeff_end}
\end{align}
and the dense output coefficients $e_i(\theta)$ for RK4-3,
\begin{align}
  e_0(\theta) &= -\nicefrac{85 \theta}{1416} \label{eq:three_step_dense_output_coeff_start}\\
  e_1(\theta) &= \nicefrac{\theta (85+2 \theta (-27+50 \theta))}{408} \\
  e_2(\theta) &= \nicefrac{\theta (131-8 \theta (24+25 \theta))}{216} \\
  e_3(\theta) &= \nicefrac{625 \theta (85+118 \theta (3+2 \theta))}{216648}.\label{eq:three_step_dense_output_coeff_end}
\end{align}

\section{Conclusion and Future Work}~\label{sec:conclusion}

In this work, we have reviewed the class of Multistep Runge-Kutta (MSRK) methods and constructed three new schemes with fourth-order accuracy. Specifically, we derived two two-step and one three-step method, with the complete set of coefficients presented in Table~\ref{tab:grand_coefficient_table}. Our stability analysis, supported by absolute stability region plots, demonstrates that the proposed MSRK schemes possess stability characteristics well-suited to partial differential equations whose spatial discretizations yield spectra concentrated along the imaginary axis.

Comprehensive stability tests using the vacuum BSSN system revealed that the RK4-2(1) and RK4-2(2) schemes can replace the standard RK4 method under identical CFL conditions typically employed in numerical relativity simulations, thereby enabling faster evolutions. We further validated these findings through a fully nonlinear binary black hole inspiral simulation using the BSSN formulation with both RK4-2 variants. Employing the $\ell^2$ norm of the Hamiltonian constraint and the extracted gravitational waves as benchmark metrics, we demonstrated that the results obtained with RK4-2 exhibit excellent agreement with those from RK4. Empirically, substituting RK4 with RK4-2 in this scenario yielded an approximate 30\% reduction in computational cost.

We additionally assessed the applicability of our methods to problems in general relativistic magnetohydrodynamics (GRMHD) using the standard Balsara 3 shock tube and Kelvin-Helmholtz instability (KHI) tests. In the Balsara 3 test, our methods produced results in close agreement with RK4. Furthermore, our methods successfully evolved the KHI instability in the presence of mesh refinement, confirming their robustness for more complex GRMHD configurations.

A comparative summary of performance across all test problems is provided in Table~\ref{tab:result_summary}. These applications collectively demonstrate the potential of MSRK methods for production-level physics simulations, particularly within the field of numerical relativity. In general, these methods offer a practical and efficient alternative to RK4 for a wide range of applications.

This study represents an initial step toward introducing MSRK methods to the numerical relativity community. We anticipate significant opportunities for further research in this area. In particular, advanced absolute stability region optimization strategies, such as those proposed in Ref.~\cite{Ketcheson_2013}, may yield even more stable and efficient methods for NR and GRMHD applications. Future work could also explore strong stability preserving (SSP) formulations and embedded MSRK variants to enable adaptive step-size control. We believe that substantial potential remains for further development and understanding of this class of methods.

\section*{Acknowledgments}

We'd like to thank Roland Haas, Helvi Witek, Cheng-Hsin Chen, and Deborah Ferguson for helpful discussions, suggestions, and comments.

JVK gratefully acknowledges financial support from CCRG through grants from NASA (Grant No. 80NSSC24K0100) and National Science Foundation (Grants No. PHY-2110338, No. PHY-2409706, No. AST-2009330, No. OAC-2031744, No. OAC-2004044).

This work was funded by NSF grants OCI 2411068, 2004157 and 2004044.

We would like to thank the Louisiana State University HPC group for their support, specifically for the use of Deep Bayou, Supermike, and Melete05.

\newpage
\appendix
\setcounter{equation}{0}
\renewcommand{\theequation}{A.\arabic{equation}}

\section{2-step method solutions}

Here we present the full solutions for the coefficient equations of the 2-step method.

\begin{align}
b_0^{(1)} & = \frac{c_2 (4-6 c_3)+4 c_3-3}{12 (c_2+1) (c_3+1)}\label{eq:two_step_sol_start}\\
b_1^{(1)} & = \frac{2 c_2 (9 c_3-5)-10 c_3+7}{12 c_2 c_3}\\
b_2^{(1)} & = \frac{7-10 c_3}{12 c_2 (c_2+1) (c_2-c_3)}\\
b_3^{(1)} & = \frac{10 c_2-7}{12 c_3 (c_3+1) (c_2-c_3)}\\
a_{20}^{(1)} & = -\frac{c_2^2}{2}\\
a_{21}^{(1)} & = \frac{1}{2} c_2 (c_2+2)\\
a_{30}^{(1)} & = \frac{c_3 \left(-2 (12 c_2+7) c_3^2-3 c_2 (5 c_2 (2 c_2+1)-4) c_3 +7 c_2 (2 c_2+3)\right)}{(6 (c_2+1)^2 (10 c_2-7))}\\
a_{31}^{(1)} & = \frac{c_3 \left(30 c_2^3 (c_3+2)+c_2^2 (4-15 c_3)+3 c_2 (c_3 (8 c_3-7)-21)+7 c_3 (2 c_3+3)\right)}{6 c_2 (c_2+1) (10 c_2-7)}\\
a_{32}^{(1)} & =  \frac{c_3 (c_2-c_3) (24 c_2 c_3+14 c_2+14 c_3+21)}{6 c_2 (c_2+1)^2 (10 c_2-7)}
\end{align}

\begin{align}
  b_0^{(2)} &= \frac{c_2 (4-6 c_3)+4 c_3-3}{12 (c_2+1) (c_3+1)}\\
  b_1^{(2)} &= \frac{2 c_2 (9 c_3-5)-10 c_3+7}{12 c_2 c_3}\\
  b_2^{(2)} &= \frac{7-10 c_3}{12 c_2 (c_2+1) (c_2-c_3)}\\
  b_3^{(2)} & = \frac{10 c_2-7}{12 c_3 (c_3+1) (c_2-c_3)}\\
  a_{20}^{(2)} &= \frac{c_2 (2 c_2 (12 c_3+7)+4 c_3 (15 c_3+8)-21)}{12 (c_3+1) (10 c_3-7)}\\
  a_{21}^{(2)} & = \frac{c_2 \left(-2 c_2 (12 c_3+7)+60 c_3^2+4 c_3-63\right)}{12 (c_3+1) (10 c_3-7)}\\
  a_{30}^{(2)} &= \frac{c_3 \left(12 (8-5 c_2) c_3^2-2 (6 c_2 (5 c_2+1)+5) c_3+7 (8 c_2-3)+120 c_3^3\right)}{12 (c_2+1) (10 c_2-7)}\\
  a_{31}^{(2)} &= \frac{c_3 \left(-120 (c_2+1) c_3^3+12 (c_2+1) (5 c_2-3) c_3^2+2 (c_2 (6 c_2 (5 c_2+1)+23)+42) c_3+c_2 (20 c_2 (6 c_2-1)-147)\right)}{12 c_2 (c_2+1) (10 c_2-7)}\\
  a_{32}^{(2)} & = -\frac{c_3 (c_3+1) (10 c_3-7) (c_2-c_3)}{c_2 (c_2+1) (10 c_2-7)}\label{eq:two_step_sol_end}
\end{align}

\section{3-Step method solutions}

Here we present the full solutions for the coefficient equations of the 3-step method.

\begin{align}
  b_0 & =\frac{10 c_3-7}{24 (c_3+2)} \label{eq:three_step_sol_start}\\
  b_1 & = \frac{11-16 c_3}{12 (c_3+1)}\\
  b_2 & = \frac{46 c_3-27}{24c_3}\\
  b_3 & = \frac{9}{4 c_3 \left(c_3^2+3 c_3+2\right)}\\
  a_{30} & = \frac{1}{12} c_3^2 (2 c_3+3)\\
  a_{31} & = \frac{1}{3} \left(-c_3^3-3 c_3^2\right)\\
  a_{32} & = \frac{c_3^3}{6}+\frac{3 c_3^2}{4}+c_3\label{eq:three_step_sol_end}
\end{align}

\section*{References}
\bibliographystyle{plain}
\bibliography{references}

\end{document}